\documentclass[prd,twocolumn,nofootinbib,aps,longbibliography]{revtex4-2}
\usepackage{amsmath,amssymb,mathtools}
\usepackage{mathrsfs}
\usepackage{tensor}
\usepackage{color,xcolor}
\usepackage{hyperref}
\hypersetup{colorlinks=true,
            linkcolor=magenta,
            filecolor=black,
            urlcolor=magenta,
            citecolor=magenta
           }
\newcommand{\smalljump}[1]{\left[\!\left[#1\right]\!\right]}
\newcommand{\bigjump}[1]{\left[\!\!\left[#1\right]\!\!\right]}
\allowdisplaybreaks
\begin{document}
\title{Tidal Love numbers of a nonexotic compact object with a thin shell}

\author{Rui-Xin Yang}
\affiliation{Department of Physics and Center for Astrophysics, Shanghai Normal University, Shanghai 200234, China}

\author{Fei Xie}
\affiliation{Department of Physics and Center for Astrophysics, Shanghai Normal University, Shanghai 200234, China}

\author{Dao-Jun Liu}
\email{djliu@shnu.edu.cn}
\affiliation{Department of Physics and Center for Astrophysics, Shanghai Normal University, Shanghai 200234, China}

\date{\today}
\begin{abstract}
As a possible alternative to black holes, horizonless compact objects have significant implications for gravitational-wave physics. In this work, we utilize the standard linearized theory of general relativity to calculate the quadrupolar tidal Love numbers of a nonexotic compact object with a thin shell proposed by Rosa and Pi\c{c}arra. It is found that both types of tidal Love numbers are positive and increase with the initial radius for almost all values of the compactness parameter. Furthermore, they have an unexpected upper bound and vanish in the most compact configurations. As a result, this model is indeed a suitable mimicker of a black hole. However, we also observed that the speed of sound within the fluid on the shell diverges in the black hole limit.
\end{abstract}
\maketitle
\section{Introduction}\label{sec:intro}
Relativistic astrophysics tells us that a sufficiently massive star cannot maintain hydrostatics equilibrium at the end of its life, and must undergo complete gravitational collapse to form a black hole (BH) \cite{Oppenheimer:1939ue}, a region where even light rays have no way to escape. While astronomers have found much evidence that BHs do exist in our universe, especially the Event Horizon Telescope Collaboration showed us the images of the supermassive BHs in the center of the galaxies \cite{EventHorizonTelescope:2019dse,EventHorizonTelescope:2022wkp}, the inevitable curvature singularity in BH spacetime is a very hot potato, at least in the purely classical theory of general relativity (GR). Observations of neutron stars have also revealed some clues that exceed the theoretical expectations \cite{LIGOScientific:2020zkf,Doroshenko2022}. These factors led to increasing interest in so-called exotic compact objects (ECOs), such as wormholes \cite{Morris:1988cz}, boson stars \cite{Schunck:2003kk}, and gravastars \cite{Mazur:2001fv}. These models can usually present a compactness arbitrarily close to that of a BH without developing any singularities in spacetime, making them potential candidates for BH alternatives. We refer readers to Ref.~\cite{Cardoso:2019rvt} for a comprehensive review.

Rosa and Pi\c{c}arra proposed two simple models of non-ECOs based on the Schwarzschild constant density star solution in Ref.~\cite{Rosa:2020hex}: the first via the collapse of the external layers of the fluid into a thin shell by performing a matching with the exterior Schwarzschild solution at a matching radius smaller than the star radius, while the second is via the creation of a vacuum bubble inside the star by matching it with an interior Minkowski spacetime. They have provided a detailed analysis of the stability and the validity of the energy condition of these two models. It is found that for a wide region of the parameter space, both models are linearly stable against radial perturbations and satisfy all of the energy conditions. More interestingly, they can present a compactness arbitrarily close to that of a BH without developing any singularities inside the objects. Recently, Rosa himself studied the observational properties of such models with accretion disks. The results indicate that the mass stored in the thin shell has a great impact on its shadow if the light ring is naked, and the most compact configuration will produce optical observational features similar to those of BHs \cite{Rosa:2023hfm}.

In gravitational-wave astronomy, the tidal Love numbers (TLNs) are a set of observable coupling constants that characterize the deformability of a self-gravitating object immersed in an external tidal field and encode the information of the object's internal structure. Flanagan and Hinderer pointed out that the tidal interaction between coalescing binary neutron stars can be measured by the current second-generation gravitational-wave detectors, and the influence of tidal effects to the waveform of an early inspiral stage is a small phase correction, which only depends on the TLNs of the neutron star \cite{Flanagan:2007ix,Hinderer:2007mb}. An intriguing fact is that the TLNs of BHs are precisely zero \cite{Binnington:2009bb}. Cardoso \textit{et al.} studied the TLNs of several ECOs and found significant differences from those of neutron stars \cite{Cardoso:2017cfl}. Thus, the presence of one of the models proposed in \cite{Rosa:2020hex} in a coalescing binary system could affect the gravitational wave signal and be potentially detectable. In this paper, we shall focus on the first model due to its relatively natural and receptive formation process.

The remainder of this paper is organized as follows. For convenience, in Sec.~\ref{sec:juncon} we briefly review the junction conditions for the metric when the spacetime features a thin shell of matter. To calculate the TLNs, in Sec.~\ref{sec:bgsol} we explicitly rewrite the metric of the first model proposed in Ref.~\cite{Rosa:2020hex}. In Sec.~\ref{sec:perconfig}, we adopt the standard linearized theory of GR to obtain the electric-type and magnetic-type TLNs of the object. Additionally, an analysis of the speed of sound on the shell is also provided. Finally, we summarize our findings with a brief discussion in Sec.~\ref{sec:conclu}.

We work in geometric units $\left(c=G=1\right)$ throughout the paper, unless otherwise noted.
\section{Junction conditions}\label{sec:juncon}
In GR, if a hypersurface $\Sigma$ separates the spacetime manifold into two regions $\mathcal{V}^{\pm}$ with different metric tensors $g_{\mu\nu}^{\pm}$, some junction conditions must be imposed on the metric to ensure that the whole spacetime is a valid distributional solution to the Einstein field equations \cite{poisson_2004}.

Let $\Sigma$ be a timelike hypersurface with matter for our purpose. Following the previous index notation conventions on the metric, we shall also label other quantities in regions $\mathcal{V}^{+}$ and $\mathcal{V}^{-}$ by attaching a superscript (or subscript) $+$ and $-$, respectively. Introducing a symbol ``$\smalljump{\quad}$'' to measure the discontinuity of a given quantity across the hypersurface, for example, $\smalljump{A}\equiv A^{+}|_{\Sigma}^{}-A^{-}|_{\Sigma}^{}$, the Darmois-Israel junction conditions can be written as \cite{MSM_1927__25__1_0,Israel:1966rt}
\begin{align}
\smalljump{\gamma_{ab}}&=0,\label{eq:1stjuncon}\\
\smalljump{K_{ab}}-\gamma_{ab}\smalljump{K}&=-8\pi S_{ab},\label{eq:2ndjuncon}
\end{align}
where $\gamma_{ab}$ and $K_{ab}$ are the first and second fundamental forms on the hypersurface, respectively, and $K$ is the trace of $K_{ab}$, while $S_{ab}$ denotes the surface stress-energy tensor that leads to the intrinsic singularity of spacetime at $\Sigma$. For the definition of $\gamma_{ab}$ and $K_{ab}$, cf., e.g., Refs. \cite{poisson_2004,Cardoso:2019upw}.

Obviously, if $S_{ab}$ vanishes, there is no matter on the hypersurface, the smooth junction conditions \eqref{eq:1stjuncon} and $\smalljump{K_{ab}}=0$ are automatically recovered.
\section{Background solution}\label{sec:bgsol}
To calculate the TLNs, we first need to explicitly get the background solution of the first model proposed in Ref.~\cite{Rosa:2020hex}. Because of the spherical symmetry of the object, we work with ordinary Schwarzschild coordinates $\{t,r,\theta,\varphi\}$.

Following Rosa and Pi\c{c}arra closely, our starting point is a Schwarzschild constant density star with mass $M$ and radius $R$. It is well-known that $R$ must be greater than $9M/4$ due to the restriction from the Buchdahl theorem \cite{Buchdahl:1959zz}; otherwise, a curvature singularity will emerge inside the star. Next, let $R_{\Sigma}$ be a radius smaller than $R$, and we leave the region of the star where $r<R_{\Sigma}$ fixed and squeeze all the matter from $R_{\Sigma}$ to $R$ onto an infinitesimally thin shell located at $R_{\Sigma}$. In general, one finally reaches a spherical object with mass $M$ and a new radius of $R_{\Sigma}$.

The surface of the object is exactly a timelike hypersurface defined by $r=R_{\Sigma}$. Following the notation conventions discussed in Sec.~\ref{sec:juncon}, hereafter, we use $+$ and $-$ to label the quantities in regions $r>R_{\Sigma}$ and $r<R_{\Sigma}$, respectively.

In this model, the spacetime geometry is governed by the metric
\begin{equation}\label{eq:bgmetric}
\overline{g}_{\mu\nu}^{\pm}=\operatorname{diag}\left[-e^{2\alpha_{\pm}(r)},e^{2\beta_{\pm}(r)},r^{2},r^{2}\sin^{2}\theta\right],
\end{equation}
with (see Appendix \ref{sec:app} for a detailed derivation)
\begin{align}
&e^{2\alpha_{+}}=e^{-2\beta_{+}}=1-\frac{2M}{r},\label{eq:extschsol}\\
&e^{2\alpha_{-}}=\left(\frac{\sqrt{1-\dfrac{2M}{R^{3}}r^{2}}-3\sqrt{1-\dfrac{2M}{R}}}{\sqrt{1-\dfrac{2M}{R^{3}}R_{\Sigma}^{2}}-3\sqrt{1-\dfrac{2M}{R}}}\right)^{\!\!\!2}\!\!\left(1-\frac{2M}{R_{\Sigma}}\right),\label{eq:bgg00}\\
&e^{-2\beta_{-}}=1-\frac{2M}{R^{3}}r^{2}.\label{eq:bgg11}
\end{align}
The solution of the ordinary constant density star is recovered when we take $R_{\Sigma}=R$, as expected.

It is straightforward to verify that the induced metrics $\gamma_{ab}$ are the same on both sides of the shell; namely, the junction condition \eqref{eq:1stjuncon} is automatically satisfied. However, the extrinsic curvature $K_{ab}$ is no longer continuous across the shell. If we regard the shell to be composed by a perfect fluid, its stress-energy tensor is simply
\begin{equation}\label{eq:surenemom}
S_{ab}=(\sigma+p_{t}^{})u_{a}u_{b}+p_{t}^{}\gamma_{ab},
\end{equation}
where $\sigma$ and $p_{t}^{}$ are, respectively, the surface energy density and the tangential pressure measured by a comoving observer $u_{a}=[-e^{\alpha(R_{\Sigma})},0,0]$. Combining Eqs. \eqref{eq:2ndjuncon}, \eqref{eq:bgmetric}, and \eqref{eq:surenemom}, it follows that
\begin{align}
\sigma&=-\frac{1}{4\pi R_{\Sigma}}\bigjump{\frac{1}{e^{\beta}}},\label{eq:bgsigma}\\
p_{t}^{}&=\frac{1}{8\pi R_{\Sigma}}\bigjump{\frac{1}{e^{\beta}}}+\frac{1}{8\pi}\bigjump{\frac{\alpha^{\prime}}{e^{\beta}}},\label{eq:bgpt}
\end{align}
where the prime denotes a derivative with respect to the radial coordinate $r$. Clearly, both $\sigma$ and $p_{t}^{}$ vanish when $R_{\Sigma}=R$.

It has been shown in \cite{Rosa:2020hex} that $\sigma>0$, $p_{t}^{}>0$, and the existence of linearly stable solutions when the weak and the strong energy conditions hold. More importantly, the junction radius $R_{\Sigma}$ can be arbitrarily close to the Schwarzschild radius $2M$ without developing any singularity inside the object. Therefore, being an exception to the Buchdahl limit, this model is a competitive BH mimicker.

For completeness, let us compare this model with the thin-shell gravastars reported in \cite{Visser:2003ge}. First, although both of them are matched to the Schwarzschild exterior through a thin shell of matter, the interior of the gravastars is a de Sitter core, while the interior of this model remains a uniform density star. Second, the surface energy density and tangential pressure of this model are both positive, but the shell of the gravastars is described by a fluid with $\sigma=0$ and $p_{t}^{}<0$. Third, the TLNs of the gravastars are negative~\cite{Cardoso:2017cfl}; however, as we shall see below, the TLNs of this model are always positive.
\section{Perturbed configuration}\label{sec:perconfig}
Imagine that a compact object introduced in the previous section is immersed in an external tidal field, which, for example, arises from its companion in a binary system. Consequently, the original spherical body is deformed by the tidal force and develops some mass multipole moments in response to the tidal field. Intuitively speaking, TLNs characterize the deformability of the objects, the bigger TLNs, the bigger deformation.

Let us consider how the spacetime geometry of the object is perturbed by the external tidal field. To do this, we will employ the standard linearized theory of GR to calculate the TLNs of the object. Then, we write
\begin{equation}\label{eq:fullmetric}
g_{\mu\nu}^{\pm}=\overline{g}_{\mu\nu}^{\pm}+h_{\mu\nu}^{\pm},
\end{equation}
where $\overline{g}_{\mu\nu}^{\pm}$ is the background metric defined by Eq.~\eqref{eq:bgmetric}, whereas $h_{\mu\nu}^{\pm}$ is a small perturbation owing to the tidal field, satisfying $|h_{\mu\nu}^{\pm}|\ll|\overline{g}_{\mu\nu}^{\pm}|$.

In the Regge-Wheeler gauge \cite{Regge:1957td}, according to the parity of the spherical harmonics $Y^{lm}(\theta,\varphi)$ under the rotation on a 2-sphere $S^{2}$, the perturbation metric can be decomposed into even and odd parts,
\begin{equation}\label{eq:permetric}
h_{\mu\nu}^{\pm}=h_{\mu\nu}^{\text{even},\pm}+h_{\mu\nu}^{\text{odd},\pm}
\end{equation}
with
\begin{align}
h_{\mu\nu}^{\text{even},\pm}&=
\begin{bmatrix}
-e^{2\alpha}H_{0}^{\pm}&H_{1}^{\pm}&0&0\\
H_{1}^{\pm}&e^{2\beta}H_{2}^{\pm}&0&0\\
0&0&r^{2}K_{\pm}&0\\
0&0&0&r^{2}\sin^{2}\theta K_{\pm}\\
\end{bmatrix}
Y^{lm},\label{eq:pereven}\\
h_{\mu\nu}^{\text{odd},\pm}&=
\begin{bmatrix}
0&0&h_{0}^{\pm}S^{lm}_{\theta}&h_{0}^{\pm}S^{lm}_{\varphi}\\
0&0&h_{1}^{\pm}S^{lm}_{\theta}&h_{1}^{\pm}S^{lm}_{\varphi}\\
h_{0}^{\pm}S^{lm}_{\theta}&h_{1}^{\pm}S^{lm}_{\theta}&0&0\\
h_{0}^{\pm}S^{lm}_{\varphi}&h_{1}^{\pm}S^{lm}_{\varphi}&0&0\\
\end{bmatrix}
,\label{eq:perodd}
\end{align}
where $H_{0},H_{1},H_{2},K,h_{0}$, and $h_{1}$ are only functions of radial coordinate $r$ and $S^{lm}_{\theta}\equiv-\partial_{\varphi}Y^{lm}/\sin\theta$ and $S^{lm}_{\varphi}\equiv\sin\theta\partial_{\theta}Y^{lm}$ are two axial vector spherical harmonics. All the functions in Eq.~\eqref{eq:permetric} are independent of time $t$ which means that the tidal field is assumed to be stationary. Actually, this is a typical scenario occurring in the inspiral stage of a coalescing binary system.

Henceforth, we shall focus our attention on the lowest quadrupolar order ($l=2$) in perturbations, because it dominates the tidal deformation of the objects. Because of the spherical symmetry of the background \eqref{eq:bgmetric}, we set the azimuthal number $m=0$ without loss of generality. For a nonrotating object, the even-parity sector decouples completely from the odd-parity sector, and we therefore can deal with them separately.
\subsection{Even-parity sector}\label{sec:even}
In this case, substituting metric \eqref{eq:fullmetric} with Eqs.~\eqref{eq:bgmetric} and \eqref{eq:pereven} into Einstein field equations and keeping only up to the first order terms in the perturbations, one obtains in turn $H_{1}^{\pm}=0$, $H_{2}^{\pm}=-H_{0}^{\pm}$, $K_{\pm}^{\prime}=-2\alpha_{\pm}^{\prime}H_{0}^{\pm}-(H_{0}^{\pm})^{\prime}$ and
\begin{align}\label{eq:K}
2e^{2\beta_{\pm}}K_{\pm}=&\left[1-3e^{2\beta_{\pm}}+r\left(\alpha_{\pm}^{\prime}+\beta_{\pm}^{\prime}-2r\alpha_{\pm}^{\prime 2}\right)\right]H_{0}^{\pm}\notag\\
&\hspace{0em}-r^{2}\alpha_{\pm}^{\prime}(H_{0}^{\pm})^{\prime}.
\end{align}
Collecting these results, by virtue of the incompressible nature of the object \cite{Damour:2009vw}, we finally obtain a single two-order homogeneous differential equation for $H_{\pm}^{}\equiv H_{0}^{\pm}$ as
\begin{equation}\label{eq:mastereqH}
H_{\pm}^{\prime\prime}+\mathcal{P}_{\pm}^{}(r)H_{\pm}^{\prime}+\mathcal{Q}_{\pm}^{}(r)H_{\pm}^{}=0,
\end{equation}
in which the coefficients are
\begin{align}
\mathcal{P}_{\pm}^{}&=\alpha_{\pm}^{\prime}-\beta_{\pm}^{\prime}+\frac{2}{r},\\
\mathcal{Q}_{\pm}^{}&=2\left(\alpha_{\pm}^{\prime\prime}-\alpha_{\pm}^{\prime}\beta_{\pm}^{\prime}-\alpha_{\pm}^{\prime 2}\right)+\frac{7\alpha_{\pm}^{\prime}+3\beta_{\pm}^{\prime}}{r}-e^{2\beta_{\pm}}\frac{6}{r^{2}}.
\end{align}

Outside the object, in the region $r>R_{\Sigma}$, Eq.~\eqref{eq:mastereqH} reduces to
\begin{equation}
H_{+}^{\prime\prime}+\frac{2\left(r-M\right)}{r\left(r-2M\right)}H_{+}^{\prime}-\frac{2\left(2M^{2}-6Mr+3r^{2}\right)}{r^{2}\left(r-2M\right)^{2}}H_{+}^{}=0,
\end{equation}
where Eq.~\eqref{eq:extschsol} was used. The general solution to the above equation in terms of the associated Legendre functions $P_{2}^{2}$ and $Q_{2}^{2}$ is found to be
\begin{equation}\label{eq:extHsol}
H_{+}^{}=a_{1}P_{2}^{2}\left(\frac{r}{M}-1\right)+a_{2}Q_{2}^{2}\left(\frac{r}{M}-1\right),
\end{equation}
where two combination coefficients $a_{1}$ and $a_{2}$ are evaluated by integrating Eq.~\eqref{eq:mastereqH} outward in the domain $r<R_{\Sigma}$ with a regular initial condition near $r=0$ and matching the exterior solution \eqref{eq:extHsol} at the surface $R_{\Sigma}$. To this end, with the help of Eqs.~\eqref{eq:fullmetric}, \eqref{eq:bgmetric}, and \eqref{eq:pereven}, we may turn the junction condition \eqref{eq:1stjuncon} at first order in the perturbations into\footnote{It should be noted that the $K$ here comes from the matrix \eqref{eq:pereven} [or Eq.~\eqref{eq:K}] instead of the trace of the extrinsic curvature in Eq.~\eqref{eq:2ndjuncon}, so be careful not to confuse them.}
\begin{equation}\label{eq:jumpHK}
\smalljump{H}=\smalljump{K}=0.
\end{equation}

Eventually, the asymptotic behavior of the solution \eqref{eq:extHsol} at infinity determines the quadrupolar electric-type TLNs as follows\footnote{The definition of $k_{l}^{\text{E,B}}$ we use in this paper is fully consistent with that of Binnington and Poisson \cite{Binnington:2009bb}, but differs from the one used by Cardoso \textit{et al.} \cite{Cardoso:2017cfl,Cardoso:2019upw} by a factor of $C^{2l+1}$.} \cite{Hinderer:2007mb,Binnington:2009bb,Damour:2009vw,Yang:2022ees}:
\begin{equation}\label{eq:k2E}
k_{2}^{\text{E}}=\frac{4}{15}C^{5}\frac{a_{2}}{a_{1}},
\end{equation}
an expression that is too complicated to be written explicitly, where $C\equiv M/R_{\Sigma}$ stands for the compactness of the object.

\begin{figure}[tbp]
\includegraphics[width=8.6cm]{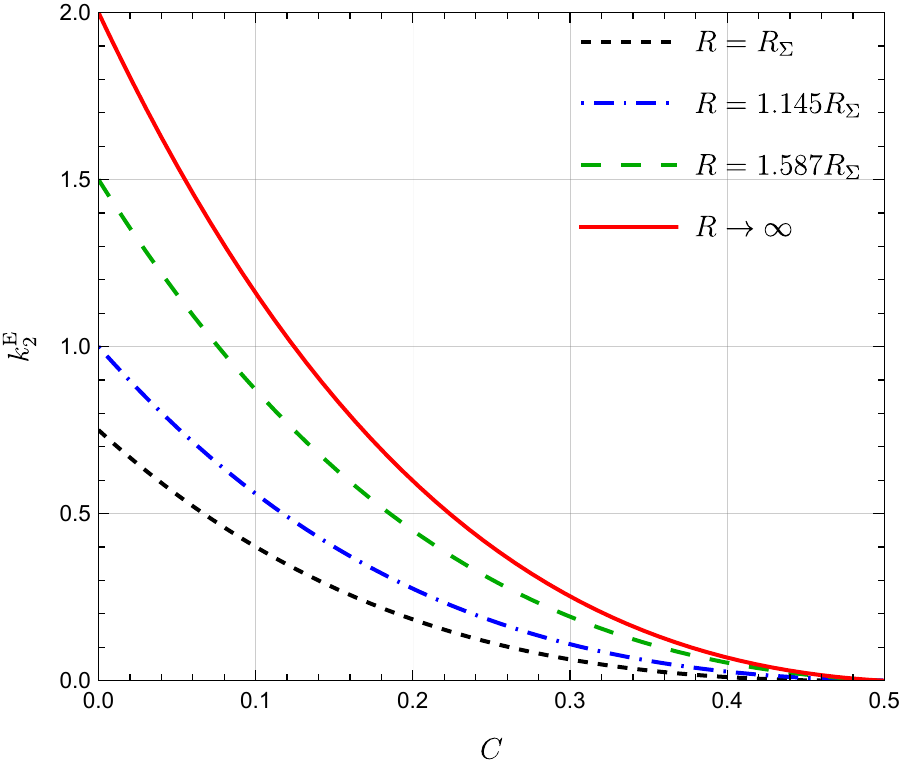}
\caption{Quadrupolar electric-type TLNs $k_{2}^{\text{E}}$ as a function of the compactness $C$ for four typical values of the initial radius~$R$. The dashed black line corresponds to a Schwarzschild constant density star and coincides with the results of Damour and Nagar \cite{Damour:2009vw}, while the solid red line corresponds to a purely thin shell of matter. All the lines vanish at the bottom right, just as a BH.}
\label{fig:k2E}
\end{figure}

The relations between the quadrupolar electric-type TLNs $k_{2}^{\text{E}}$ and the compactness $C$ of the object are illustrated in Fig.~\ref{fig:k2E}, and those typical values of the initial radius $R$ are obtained in the Newtonian limit ($C\to 0$) of Eq.~\eqref{eq:k2E}. When $R=R_{\Sigma}$, in the absence of the shells, we recover the profile of Schwarzschild constant density stars \cite{Damour:2009vw}, as expected. Regardless of the value of the initial radius $R$, we identify that the $k_{2}^{\text{E}}$ vanishes in the BH limit ($C\to 0.5$). Besides this extreme case, for each certain equilibrium configuration, the values of $k_{2}^{\text{E}}$ are monotonically increasing with the increase in the initial radius $R$. However, the $k_{2}^{\text{E}}$ cannot be arbitrarily large, as it has an upper bound derived from the limit of $R\to\infty$, i.e.,
\begin{equation}
k_{2}^{\text{E}}=\frac{8}{5}C^{5}\frac{\xi(\xi+1)}{4C(C^{2}-6C+3)+3\xi(\xi+1)\ln\xi}
\end{equation}
with $\xi\equiv 1-2C$. This is nothing but a purely thin shell of matter.
\subsection{Speed of sound on the shell}
If we do not consider the thin-shell material to be stiff and supplement an equation of state that relates the surface density to pressure,
\begin{equation}
p_{t}^{}=p_{t}^{}(\sigma),
\end{equation}
then the speed of sound within the fluid is well-defined and can be obtained reasonably.

Taking into account that $H$ itself is a continuous function across the shell, the junction condition \eqref{eq:2ndjuncon} along with Eqs.~\eqref{eq:fullmetric},~\eqref{eq:bgmetric}~and~\eqref{eq:pereven}, to first order in the perturbations, implies that
\begin{align}
&8\pi\delta\sigma=\bigjump{\frac{H^{\prime}}{e^{\beta}}}-\frac{H(R_{\Sigma})}{R_{\Sigma}}\bigjump{\frac{1}{e^{\beta}}}+2H(R_{\Sigma})\bigjump{\frac{\alpha^{\prime}}{e^{\beta}}},\\
&16\pi\delta p_{t}^{}=\frac{H(R_{\Sigma})}{R_{\Sigma}}\bigjump{\frac{1}{e^{\beta}}}-H(R_{\Sigma})\bigjump{\frac{\alpha^{\prime}}{e^{\beta}}},
\end{align}
where $\delta\sigma$ and $\delta p_{t}^{}$ are the fluctuations of thin-shell matter to the background \eqref{eq:bgsigma} and \eqref{eq:bgpt}, respectively. Hence, the speed of sound in the fluid is evaluated by the adiabatic relation
\begin{equation}
v_{s}^{2}=\frac{\delta p_{t}^{}}{\delta\sigma},
\end{equation}
which depends on $C$, $R$, $H_{-}^{}(R_{\Sigma})$, and $H_{-}^{\prime}(R_{\Sigma})$.

\begin{figure}[htbp]
\includegraphics[width=8.6cm]{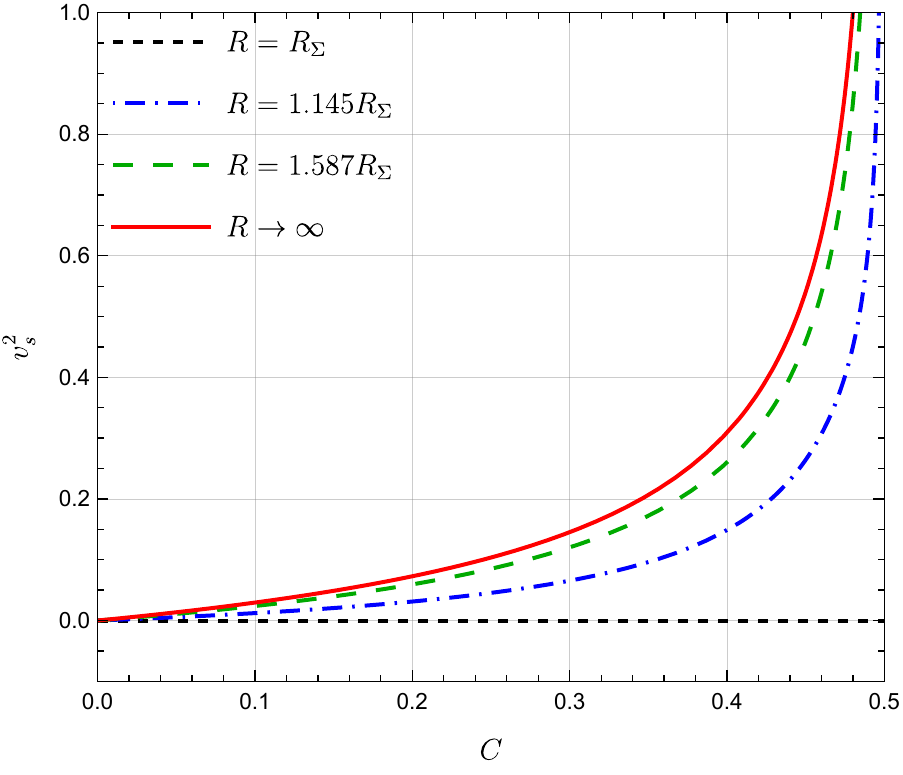}
\caption{Speed of sound $v_{s}^{2}$ on the shell against the compactness $C$ for a few values of $R$. Except for the case of $R=R_{\Sigma}$ (dashed black curve), it is easily found that $v_{s}^{2}$ ranges between $(0,1)$ for a wide region of $C$, but diverges in the BH limit. Actually, the sound speed of the $R\to\infty$ configuration (solid red curve) has already reached the speed of light, when $C=0.48$ \cite{Frauendiener_1990,Andreasson:2007ck}.}
\label{fig:vs2}
\end{figure}

For nonexotic matter, the magnitude of sound speed should fall within the interval $0<v_{s}^{2}<1$. Indeed, for a wide region of the compactness parameter $C$, it is easily verified that the $v_{s}^{2}$ on the shell is positive and less than one everywhere, as long as $R\neq R_{\Sigma}$. The speed of sound is identically vanishing in the case of $R=R_{\Sigma}$, which is caused by the density step at the surface of the Schwarzschild constant density stars. Nevertheless, the values of $v_{s}^{2}$ increase with increasing compactness $C$, and ultimately tend to infinity in the BH limit; see Fig.~\ref{fig:vs2}.
\subsection{Odd-parity sector}
For odd-parity perturbations \eqref{eq:perodd}, the linearized Einstein field equations require $h_{1}^{\pm}=0$ and $h_{\pm}^{}\equiv h_{0}^{\pm}$ satisfying the following equation:
\begin{equation}\label{eq:mastereqh}
h_{\pm}^{\prime\prime}+\mathcal{F}_{\pm}^{}(r)h_{\pm}^{\prime}+\mathcal{G}_{\pm}^{}(r)h_{\pm}^{}=0
\end{equation}
with
\begin{align}
\mathcal{F}_{\pm}^{}&=-\left(\alpha_{\pm}^{\prime}+\beta_{\pm}^{\prime}\right),\\
\mathcal{G}_{\pm}^{}&=-2\left(\alpha_{\pm}^{\prime\prime}-\alpha_{\pm}^{\prime}\beta_{\pm}^{\prime}+\alpha_{\pm}^{\prime 2}\right)-e^{2\beta_{\pm}}\frac{6}{r^{2}}.
\end{align}

Outside the object, the above equation reduces to
\begin{equation}
h_{+}^{\prime\prime}+\frac{2\left(2M-3r\right)}{r^{2}\left(r-2M\right)}h_{+}^{}=0,
\end{equation}
and the general solution of this simple equation can be expressed as \cite{Abdelsalhin:2019ryu,Yang:2022ees}
\begin{align}\label{eq:exthsol}
h_{+}^{}&=b_{1}\left(\frac{r}{2M}\right)^{\!2}{ }_{2}F_{1}\left(-1,4,4;\frac{r}{2M}\right)\notag\\
&\hspace{1.2em}+b_{2}\left(\frac{2M}{r}\right)^{\!\!2}{ }_{2}F_{1}\left(1,4,6;\frac{2M}{r}\right),
\end{align}
where $_{2}F_{1}$ denotes the hypergeometric function, while constants $b_{1}$ and $b_{2}$ are determined by the interior numerical solution of master Eq.~\eqref{eq:mastereqh}, exterior solution~\eqref{eq:exthsol}, and the following junction conditions:
\begin{gather}
\smalljump{h}=0,\\
\bigjump{\frac{h^{\prime}}{e^{\beta}}}=2h(R_{\Sigma})\bigjump{\frac{\alpha^{\prime}}{e^{\beta}}},
\end{gather}
which are derived by plugging Eqs.~\eqref{eq:fullmetric}, \eqref{eq:bgmetric}, and \eqref{eq:perodd} into Eqs.~\eqref{eq:1stjuncon} and \eqref{eq:2ndjuncon}, respectively.

Finally, the coefficients $b_{1}$ and $b_{2}$ are related to the quadrupolar magnetic-type TLNs via \cite{Binnington:2009bb,Yang:2022ees}
\begin{equation}\label{eq:k2B}
k_{2}^{\text{B}}=-\frac{32}{3}C^{5}\frac{b_{2}}{b_{1}}.
\end{equation}

\begin{figure}[htbp]
\includegraphics[width=8.6cm]{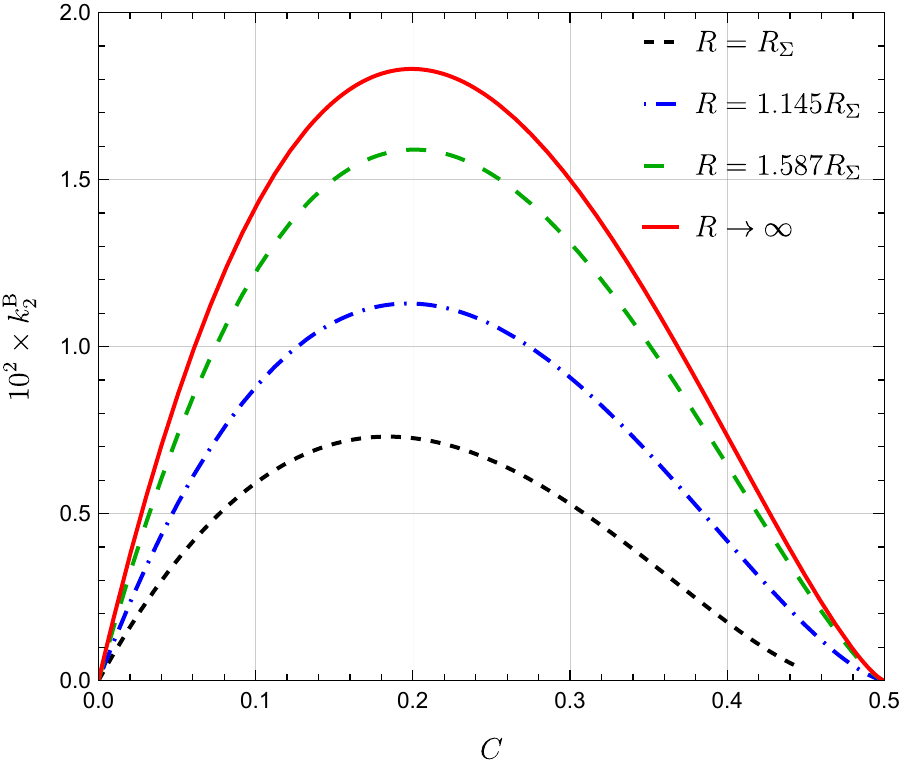}
\caption{The 100 times quadrupolar magnetic-type TLNs $k_{2}^{\text{B}}$ versus the compactness $C$ for different values of $R$.}
\label{fig:k2B}
\end{figure}

In Fig.~\ref{fig:k2B}, we plot the quadrupolar magnetic-type TLNs $k_{2}^{\text{B}}$ as a function of the compactness $C$ for the same typical values of $R$. Similar to $k_{2}^{\text{E}}$, $k_{2}^{\text{B}}$ is also positive, increases with the initial radius $R$ for a given compactness $C$, and possesses an upper bound. However,  $k_{2}^{\text{B}}$ is not a monotonical function of $C$ for a given value of $R$. In the BH limit, $k_{2}^{\text{B}}$ vanishes. Meanwhile, $k_{2}^{\text{B}}$ also converges to zero in the Newtonian limit due to the fact that $k_{2}^{\text{B}}$ is fully relativistic.
\section{Conclusions and discussions}\label{sec:conclu}
In the present work, we have considered a class of compact objects proposed by Rosa and Pi\c{c}arra \cite{Rosa:2020hex} that consist of a constant density star with some of its mass collapsed into a thin shell. The freedom to choose the radius of the compact object in such a model means that it can be arbitrarily close to the Schwarzschild radius, thus providing a model of compact objects that could mimic a BH. We computed both  the electric-type and magnetic-type quadrupolar TLNs of the objects. The results show that the matter stored in the thin shell has a significant effect on the TLNs. It is found that both types of TLNs cannot be arbitrarily raised, but are bounded by the property of matter on the thin shell. Furthermore, for almost all final equilibrium configurations of the object, both types of TLNs increase as the initial radius increases, and regardless of the value of the initial radius, they both vanish in the most compact configurations of the object. Therefore, this nonexotic compact object with a thin shell is indeed an exception to the Buchdahl theorem and does serve as a suitable BH mimicker.

It is remarkable that, compared with those of ECOs such as wormholes and  thin-shell gravastars, which are negative in the case of high compactness \cite{Cardoso:2017cfl}, both types of TLNs of the object are always positive. Note that the TLNs of the object are also different from those of boson stars in this respect, because the magnetic-type TLNs of boson stars are negative, even though the electric-type are positive. 

Given the sizable value of the TLNs of the object, it is expected that current gravitational-wave detectors will be able to constrain the parameters of the object. However, although there exist differences between the tidal deformability of this object with those of BHs in modified theories of gravity and of other ECOs, it is difficult to distinguish the object from BHs and other ECOs by measuring its TLNs via gravitational-wave observations, especially in the high-compactness regime.  The situation may change as the next-generation ground-based detectors such as ET \cite{Maggiore_2020} or space-based detectors such as LISA \cite{LISA:2017pwj} are  put into operation.

In addition, we conducted an analysis of the speed of sound within the fluid on the shell using a unified framework. For a wide region of the compactness parameter of the model, we verify that the speed of sound is less than the speed of light. Thus, the shells are composed by physically relevant matter in these circumstances. However, the sound speed on the shell diverges in the BH limit, which means that the thin-shell matter becomes increasingly stiff as the compactness increases.

Finally, it would be interesting to investigate whether the I-Love-Q relations \cite{Yagi:2013awa} remain valid by anchoring an angular momentum onto this model. This issue, however, extends from the scopes of the present work.
\begin{acknowledgments}
The authors thank Xing-Hua Jin, Yu Wang, and Zhou-Li Ye for helpful discussions. This work is supported by the innovation program of Shanghai Normal University under Grant No.~KF202147.
\end{acknowledgments}
\appendix
\section{DERIVATION OF THE BACKGROUND METRIC}\label{sec:app}
To avoid confusion, we here provide a derivation of the background metric in Sec.~\ref{sec:bgsol}.

First, the spacetime outside the object is definitely described by the exterior Schwarzschild solution \eqref{eq:extschsol}. Then, we assume that the matter inside the object can be modeled as an isotropic perfect fluid, and its energy-momentum tensor is given by
\begin{equation}\label{eq:bgenemom}
\tensor{T}{_\mu^\nu}=\operatorname{diag}\left[-\rho(r),p(r),p(r),p(r)\right],
\end{equation}
where $\rho$ is the energy density and $p$ the pressure. The Einstein field equations under the ansatz \eqref{eq:bgmetric} and \eqref{eq:bgenemom} yield
\begin{align}
e^{-2\beta_{-}}&=1-\frac{2m}{r},\label{eq:EFE00}\\
\alpha_{-}^{\prime}&=\frac{4\pi p r^{3}+m}{r(r-2m)},\label{eq:EFE11}\\
p^{\prime}&=-(\rho+p)\alpha_{-}^{\prime},\label{eq:EFE22}
\end{align}
where $m$ is the mass function defined by
\begin{equation}\label{eq:masfun}
m(r)\equiv 4\pi\int_{0}^{r}\rho x^{2}\,dx.
\end{equation}

Note that in the model we are considering, only the external layers of a constant density star are collapsed, and nothing happened in the region $r<R_{\Sigma}$, which means that neither the density nor the central pressure of the object has changed~\cite{Rosa:2020hex},
\begin{equation}
\rho=\frac{3}{4\pi}\frac{M}{R^{3}},\quad p(0)=\rho\frac{\sqrt{1-\dfrac{2M}{R}}-1}{1-3\sqrt{1-\dfrac{2M}{R}}}.
\end{equation}
Thus, from Eqs.~\eqref{eq:EFE00}-\eqref{eq:masfun} one obtains Eq.~\eqref{eq:bgg11}, 
\begin{equation}
p=\rho\frac{\sqrt{1-\dfrac{2M}{R}}-\sqrt{1-\dfrac{2M}{R^{3}}r^{2}}}{\sqrt{1-\dfrac{2M}{R^{3}}r^{2}}-3\sqrt{1-\dfrac{2M}{R}}},
\end{equation}
and 
\begin{equation}
e^{2\alpha_{-}}=\frac{c_{1}}{(\rho+p)^{2}}\label{eq:c1},
\end{equation}
where $c_{1}$ is an integration constant to be determined by the junction condition \eqref{eq:1stjuncon} with Eq.~\eqref{eq:bgmetric}, i.e.,
\begin{equation}
e^{2\alpha_{-}(R_{\Sigma})}=e^{2\alpha_{+}(R_{\Sigma})}=1-\frac{2M}{R_{\Sigma}}.
\end{equation}
Using these expressions above to eliminate $c_{1}$ from Eq. \eqref{eq:c1}, we finally arrive at Eq. \eqref{eq:bgg00}.
\bibliography{ref}
\end{document}